\documentclass[preprint,12pt]{elsarticle}

\usepackage{amssymb}
\usepackage{amsmath}
\usepackage{hyperref}
\hypersetup{
     colorlinks = true,
     citecolor  = cyan,  
     linkcolor  = magenta 
}

\usepackage{subcaption}

\journal{Physics Letters A}

\begin{document}

\begin{frontmatter}



\title{Towards an understanding of dipole-dipole interactions in nonlocal media}



  \author[label1]{L. In\'acio}
    \author[label1]{A. Kurumbail}
      \author[label1,label4,label5]{S. K. Panja}
            \author[label2]{I. Brevik}
  \author[label1,label3]{M. Bostr\"om\corref{cor}}
\ead{mathias.bostrom@ensemble3.eu}
 
\cortext[cor]{Corresponding author} 
 \affiliation[label1]{organization={Centre of Excellence ENSEMBLE3},
             addressline={Wolczynska Str. 133},
             city={Warsaw},
             postcode={01-919},
             country={Poland}}

\affiliation[label4]{organization={Wilczek Quantum Center, Shanghai Institute for Advanced Studies},
city={Shanghai}, postcode={201315},
             country={China}}

\affiliation[label5]{organization={University of Science and Technology of China},
city={ Hefei}, postcode={230026},
             country={ China}}
             
\affiliation[label2]{organization={Department of Energy and Process Engineering},
             city={Trondheim},
             postcode={NO-7491},
             country={Norway}}
\affiliation[label3]{organization={Chemical and Biological Systems Simulation Lab, Centre of New Technologies, University of Warsaw},
addressline={Banacha 2C},
city={Warsaw},
postcode={02-097},
country={Poland}}

\begin{abstract}
We commence our study with review of dispersion interactions in electrolytes. 
We then reflect on how background media change atom-atom excited-state systems. To highlight the impact of nonlocal media, such as salt solutions, we predict that a new contribution to the resonance interaction energy emerges in a form $\propto e^{-\kappa_{\rm D} \rho}/\rho$. Here $\kappa_{\rm D}$ is the Debye length and $\rho$ is the distance between the atoms. This contribution vanishes at zero temperature, where a new term proportional to $1/\rho^4$ (similar to free space) occurs. This new term is dampened by the electrolyte at large distances, causing it to decrease much faster, proportional to $1/\rho^7$. The long-range electrolyte-induced resonance interaction at finite temperature may, in addition to the dominating van der Waals attraction (which goes as $1/\rho^6$), take part in the molecular formation of biological fluids.
\end{abstract}

\begin{graphicalabstract}
\includegraphics[width=\columnwidth]{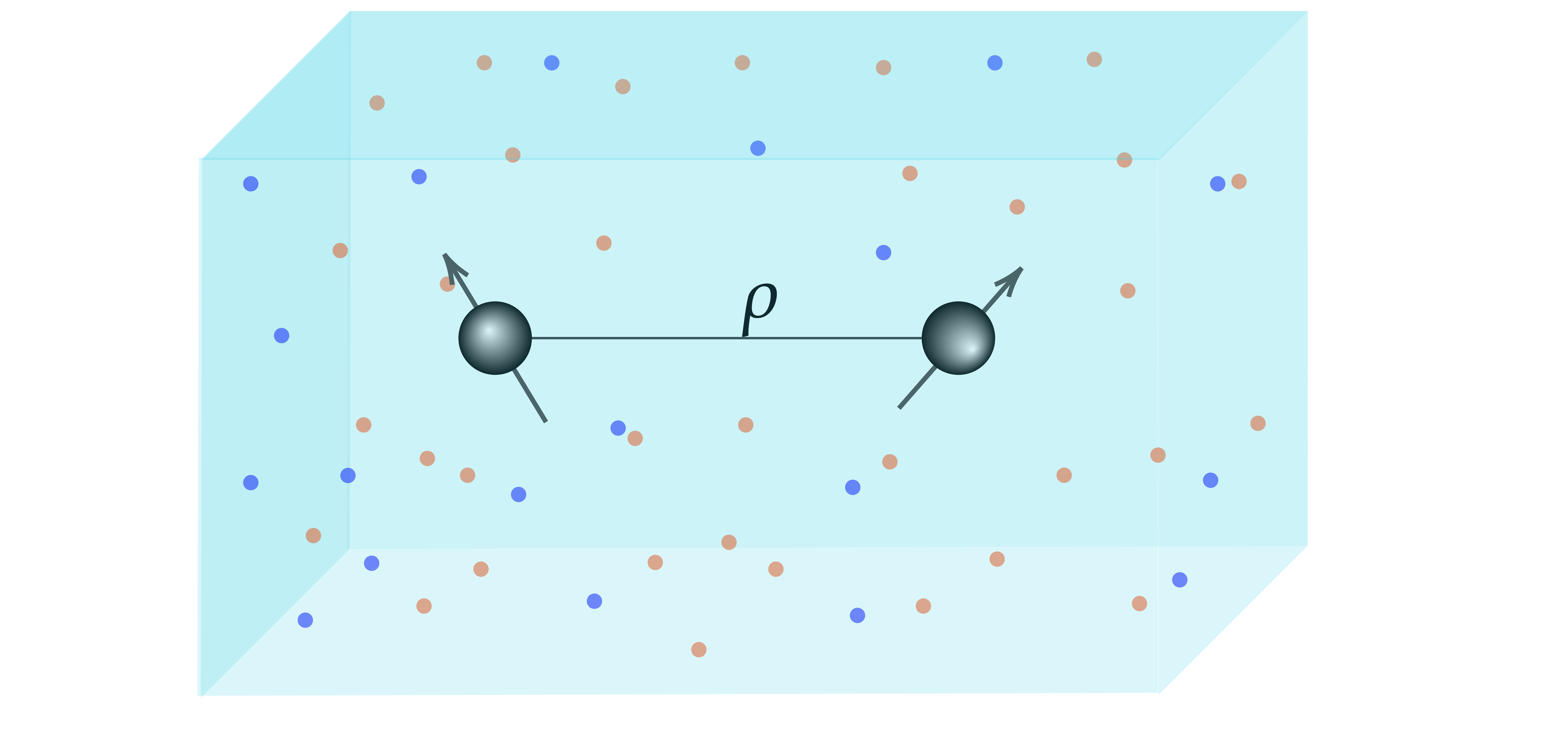}
\end{graphicalabstract}

\begin{highlights}
\item Semi-classical electrodynamics theory for atom-atom interaction
\item Resonance interaction theory revisited and expanded 
\item Non-local effects on excited state interactions
\item Proposed model for pheromone action via energy transfer
\end{highlights}


\end{frontmatter}





\section{Introduction}

\begin{figure}[!h]
    \centering
    \includegraphics[width=0.7\linewidth]{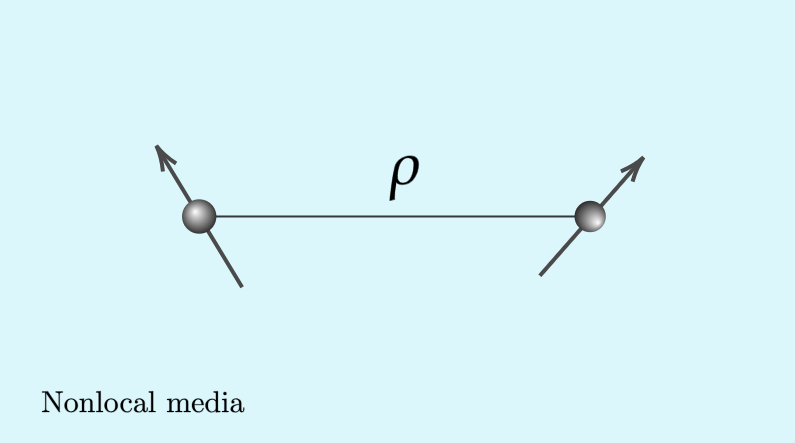}
    \caption{Color online: Schematic figure for two atoms in an excited state, a distance $\rho$ apart, in a nonlocal media. }
    \label{fig:scheme}
\end{figure}

Many biological processes occur in conductive media, where dispersion interactions\,\cite{Munday2009,Zwol1,RevModPhys.88.045003} are modified relative to free space.
Nonlocal effects alter the van der Waals forces between surfaces\,\cite{MaiaNeto2019,nunes2021casimir,PhysRevA.111.012816}, as the movement of ions affects the medium's response. 
Similarly, we will demonstrate that the presence of ions also influences the resonance interaction between two particles in solution (e.g., in extensions of the present work at the boundary between a complex protein and electrolyte). Such non-local effects in salt solutions may even impact cell physiology. Our model system, consisting of two atoms in an excited state in a nonlocal medium, is shown in Fig.\,\ref{fig:scheme}.

We derive expressions for salt-modified excited-state resonance van der Waals interactions between atom pairs in an orientationally averaged excited state when immersed in an electrolyte. Past work indicated\,\cite{Bostrom1,BostromPRA} that because of drastic approximations, the underlying theory of resonance interactions in free space derived from perturbative quantum electrodynamics (QED) may need to be revised in both the long-range retarded and high temperature/large separation entropic limits. To demonstrate how this impacts non-local interactions in fluids, we use the classical theory for van der Waals potentials between ground-state atom pairs in an electrolyte\,\cite{Maha}.

\section{Background to Dispersion Forces in Biology}

The focus of the current work is to discuss how ions screen non-local dispersion potentials in salt solutions. We will, in particular, explore excited-state dispersion potentials as an example where the interaction is modified due to background media.
F\"{o}rster energy transfer (F\"oster Resonant Energy Transfer - commonly called FRET or RET) uses a dipole-dipole mechanism\,\cite{Forster} and is crucial in understanding many mechanisms in biophysics\,\cite{govorov2017understanding}. It underlies photosynthesis, artificial light harvesting, and fluorescent light-emitting devices\,\cite{Groendelle1}. FRET was discovered experimentally by Cario and Franck in 1923\,\cite{Cario}, but questions remain regarding its properties and relevant effects due to confinement, solvent, and temperature. The dipole-dipole mechanism has furthermore been proposed as a way to create entangled states for quantum logic using molecules\,\cite{Brennen}. However, while the mechanism certainly is correct, past work has indicated that the underlying theory derived from perturbative quantum electrodynamics (QED) may need to be revised in the retarded and entropic limits\,\cite{Bostrom1}. A result that appears to be right—for systems at zero temperature -- was found by Ninham and his group more than 50 years ago. This was not made public until 2003\,\cite{Bostrom1}, where the theory was extended to account for finite temperatures. In the current work, we describe non-local resonance interaction in the presence of an electrolyte with a low salt concentration.

\section{The Semi-Classical Approach to Atomic Dispersion Forces}

To illustrate the power of the original derivation\,\cite{Bostrom1}, before we go beyond past work and model non-local effects, consider a system with two identical molecules—one in its ground state and the other in an excited state. This can also be represented by a superposition: symmetric and antisymmetric states regarding the interchange of the molecules. Although the symmetric case is likely to decay into two ground-state molecules, the antisymmetric case can be quite long-lived, leading to the system being trapped\,\cite{Stephen}. The energy migrates back and forth between the two molecules until they either move apart or the system emits a photon. This coupling causes first-order dispersion interactions; the energy difference between the two states depends on the separation $\rho$. The resonance interaction energy is then used to describe the energy transfer rate. 
In the weakly coupled and non-retarded limit, this reproduces the F\"{o}rster energy transfer rate. However, the results\,\cite{Bostrom1} are qualitatively different in the retarded regime when compared to previous derivations. The correct asymptote can only be obtained when one accounts for finite temperature.

 Recall a well-known system\,\cite{McLachlan} consisting of two identical polarizable dipoles, denoted 1 and 2, with a coordinate system placing the origin at dipole 1 and the axis directed toward dipole 2.\,\cite{Maha,Bostrom1} The oscillating moment of the dipoles 
 gives rise to an oscillating field 

\begin{equation}
\mathbf{E}(\mathbf{\rho} \mid i)=\mathbf{T}(\mathbf{\rho}, \omega) \mathbf{P}(i),
\end{equation}
where $\textbf{E}(\rho\mid i)$ is the electric field strength of dipole $i$ in the position $\rho$, $\textbf{P}(i)$ is the oscillating strength of the $i$-dipole, and $\textbf{T}(\rho,\omega)$ is the tensor related to the susceptibility\cite{Maha,Bostrom1} 
\begin{equation}
\mathbf{T}(\mathbf{\rho}, \omega)_{i j}=\left(\frac{\partial}{\partial x_{i}} \frac{\partial}{\partial x_{j}}-\delta_{i j} \nabla^{2}\right) G(\mathbf{\rho}, \omega).
\end{equation}

The Green's function in this case is\,\cite{Maha,Bostrom1}
\begin{equation}
G(\mathbf{\rho}, \omega )=\frac{1}{(2 \pi)^{3}} \int \mathrm{~d}^{3} k \, \frac{\mathrm{e}^{\mathrm{i} \mathbf{k} \mathbf{\rho}}}{\omega^{2} / c^{2}-k^{2}}\,,
\end{equation}
and it yields the following non-zero matrix elements\,\cite{Maha,Bostrom1}

\begin{align}
T_{x x}=T_{y y} & =\left(\frac{\omega^{2}}{c^{2}}+\frac{\mathrm{i} \omega}{\rho c}-\frac{1}{\rho^{2}}\right) \frac{\mathrm{e}^{\mathrm{i} \omega \rho / c}}{\rho}, \\
T_{z z} & =2\left(\frac{1}{\rho^{2}}-\frac{\mathrm{i} \omega}{\rho c}\right) \frac{\mathrm{e}^{\mathrm{i} \omega \rho / c}}{\rho}. 
\end{align}
The matrix \textbf{T} can then be separated into real and imaginary parts\,\cite{Maha,Bostrom1}
\begin{equation}
\mathbf{T}(\mathbf{\rho}, \omega)=\mathbf{T}^{\prime}(\mathbf{\rho}, \omega)+\mathrm{i} \mathbf{T}^{\prime \prime}(\mathbf{\rho}, \omega),
\end{equation}
which give rise to both in-phase and out-of-phase electric field components. In the limit $|\mathbf{\rho}| \rightarrow 0$, the real part of $G$ is a principal value integral identically zero for all real $\omega$. The imaginary part yields an out-of-phase contribution to $\mathbf{T}(\rho,\omega)$, which is the well-known radiation damping\,\cite{Bostrom1}
\begin{equation}
\mathbf{T}^{\prime \prime}(0, \omega)_{i j}= \frac{2}{3} \frac{\omega^{3}}{c^{3}} \delta_{i j} \equiv \frac{\omega}{\alpha_{0}} \gamma(\omega)\delta_{i j} .
\end{equation}
Here $\gamma(\omega)$ is the damping strength and $\alpha_0$ is the isotropic polarizability. Thus, the equation of motion for two coupled dipoles separated by a distance $\rho$ is\,\cite{McLachlan,Maha,Bostrom1}
\begin{equation}
    \left(\omega_{0}^{2}-\omega^{2}\right) \mathbf{P}(1)=\alpha_{0}\left[\mathrm{i} \mathbf{T}^{\prime \prime}(0, \omega) \mathbf{P}(1)+\mathbf{T}(\mathbf{\rho}, \omega) \mathbf{P}(2)\right],
\end{equation}
\begin{equation}
    \left(\omega_{0}^{2}-\omega^{2}\right) \mathbf{P}(2)=\alpha_{0}\left[\mathrm{i} \mathbf{T}^{\prime \prime}(0, \omega) \mathbf{P}(2)+\mathbf{T}(\mathbf{\rho}, \omega) \mathbf{P}(1)\right]. 
\end{equation}
Here $\alpha_{0}=f e^{2} / m$, and $f$ is the strength, $e$ the unit electric charge, $m$ the mass, and $\omega_{0}$ the frequency of the oscillator. The isotropic polarizability $\alpha_0$ is related to the polarizability tensor as

\begin{align}
    {\boldsymbol\alpha}_{ij}(\omega)=\frac{\alpha_0}{\omega_0^2-\omega^2-i\omega\gamma(\omega)}{\boldsymbol \delta_{ij}}={\alpha}(\omega){\boldsymbol\delta}_{ij}\,.
\end{align}

Finally, we can connect the equations for each dipole in
\begin{align}
    \textbf{P}(1)=\alpha(\omega)\textbf{T}(\rho,\omega)\textbf{P}(2)\, ,\\
    \textbf{P}(2)=\alpha(\omega)\textbf{T}(\rho,\omega)\textbf{P}(1)\, . 
\end{align}
This treatment\,\cite{Maha,Bostrom1} generally agrees with alternative formulations for ground and excited state descriptions done via perturbative quantum field theory in certain conditions. This indicates that perturbative quantum electrodynamics should be further tested at large separations and finite temperatures. A good starting point should originate from the two books by Buhmann\,\cite{Buhmann12a,Buhmann12b}.

\section{Reflecting on Potential Flaws in Perturbative QED}

We recapitulate\,\cite{Bostrom1} in this section how the excited state dispersion interaction in free space\,\cite{Avery,AndrewsCurutchetScholes2011}, derived from perturbative QED, can be re-obtained in the semi-classical formalism after performing specific approximations\,\cite{Bostrom1}. The pole of the polarizability, i.e., the resonance frequency, is close to the oscillator frequency $\omega_j$. Hence, a common method is to replace the field susceptibility matrix $\textbf{T}(\rho,\omega)$ with $T(\rho,\omega_j)$, and drop the dissipation frequency (i.e., the lifetime)\cite{McLachlan,Brennen}. The resonance frequency is within the approximation
\begin{equation}
\omega_r\approx\omega_j \sqrt{1+\alpha_0 T(\rho,\omega_j)}\approx \omega_j+\frac{\omega_j \alpha_0 T(\rho,\omega_j)}{2}\, .
\label{Eq7}
\end{equation} 
Using definitions of oscillator strength and static polarizability, we can write the interaction energy as\cite{McLachlan}
 \begin{equation}
U(\rho)=p^2 T(\rho,\omega_j),
\label{Eq8}
\end{equation}
where $p$ is the magnitude of the transition dipole moment. This is a classic textbook result obtained in various derivations over 60 years. Another way to derive the same result is to first calculate the displacement vector field of the non-interacting excited molecule and then calculate the energy of the dipole of the other molecule within this field \cite{Power,Power2}. 
The coupling of the system was ignored in the earlier derivations performed prior to the work by Ninham and co-workers\,\cite{Bostrom1}. In the strong-coupling limit, the retarded asymptotic transfer rate is $n \propto \rho^{-1} cos(\omega_j \rho/c)$, whereas in the weak-coupling limit the transition rate is, as before, obtained using the Fermi golden rule. The real part of the field susceptibility is used in the standard approach \cite{Andrews}. The averaged isotropic retarded transfer rate becomes proportional to $n \propto cos^2(\omega_j \rho/c)/\rho^2$. However, the oscillating transfer has been recognized as incorrect; the complex field susceptibility must be used to avoid these oscillations. After averaging isotropically, the following result is obtained \cite{Avery,Power,Power2,Andrews}
\begin{equation}
n \propto \frac{3}{\rho^6}+\bigg(\frac{\omega_j}{c}\bigg)^2\frac{1}{\rho^4}+\bigg(\frac{\omega_j}{c}\bigg)^4\frac{1}{\rho^2}.
\end{equation}
In the non-retarded limit, this result is identical to what was derived in the semi-classical approach\,\cite{Bostrom1}, but this changes completely in the retarded limit. The argument typically used to support the dependence proportional to $\rho^{-2}$ is that it supposedly corresponds to real photon exchange and is characteristic of a classical spherical wave. This separation dependence also results in unphysical infinities for an increasingly large system of molecules. A way to avoid these is to consider the influence of the background medium. In this way, an exponentially decaying factor modulates the interaction in condensed matter. However, this approach and its underlying theory are in great need of further exploration. In the next section, we will explore how non-perturbative theory predicts terms that depend on Debye screening when two atoms in an excited state interact within an electrolyte.

\section{Theory For Non-Local First Order Dispersion Forces}

Recall our earlier system of two identical atoms—one in ground state and one excited—with first-order dispersion interactions caused by coupling. Before we initiate our exploration on the impact of non-local effects, we note that the presence of background media will cause the excited states to be short-lived and impact real photon exchange.

After writing down the equations of motion for the excited system, it is straightforward to derive the zero-temperature Green's function for each of two atoms \,\cite{McLachlan,Bostrom1}. The resonance frequencies of the system are given by the following equation\,\cite{McLachlan}

\begin{equation}
1-{\boldsymbol\alpha}^{\rm (1)}(\omega)^* \, {\boldsymbol\alpha}^{\rm (2)}(\omega)^* \, \textbf{T}(\rho|\omega)^2=0.
\label{Eq1}
\end{equation}

Here ${\boldsymbol\alpha}^{\rm (i)}(\omega)^*$ is the excess polarizability of the atom $i$ (here $i=1, 2$) in salt solution. This resonance condition can be separated into antisymmetric and symmetric parts, which can then be used to calculate the ground state van der Waals interaction between particles in solutions\,\cite{Buhmann12a,Buhmann12b}.

Since the excited symmetric state is assumed to have a much shorter lifetime than the antisymmetric one, the system can be trapped in the latter\,\cite{Bostrom1}.  
The resonance interaction energy of this is
\begin{equation}
\label{Eq3}
U(\rho)= \hbar [\omega_{r} (\rho)-\omega_{r} (\infty)].
\end{equation}
Since the relevant solution of q.\,(\ref{Eq1}) is really the pole associated with the antisymmetric part of the underlying Green's function, we can, in a standard way\,\cite{Maha}, deform a contour of integration around this pole to obtain a simple and exact expression for the resonance interaction energy
\begin{eqnarray}
U(\rho) = (\hbar/ \pi) \int_0^\infty d \xi \ln[1+{\boldsymbol\alpha}^{\rm (1)}(i \xi)^* \textbf{T}(\rho, i \xi)].
\label{Eq4}
\end{eqnarray}
To account for the temperature $T$ dependence, we replace the integration over imaginary frequencies with a summation over discrete frequencies\,\cite{Maha}
\begin{equation}
{\frac{\hbar}{2 \pi}} \int_0^\infty d \xi \rightarrow k_B T
\sum_{n=0}^{\infty}{'},\quad \xi_n=2 \pi n  \frac{k_B T}{\hbar},
\label{Eq5}
\end{equation}
where $k_B$ is the Boltzmann constant and the prime indicates that the $n=0$ term should be halved.

 For the usual concentrations of biological electrolytes (e.g., a salt concentration of 0.1 M), the ionic plasma frequency is many orders of magnitude smaller than the first non-zero Matsubara frequency \footnote{The first Matsubara frequency is $\propto 10^{14}$ rad/s and the characteristic plasma frequency for such electrolyte is $\sim 10^{12}$ rad/s.}. The plasma frequency is even smaller for pure water at pH 7. At higher frequencies, the ionic plasma will have relaxed. 
 Therefore, in both free space and in electrolyte, the field susceptibility matrix, ${\bf  T}(\rho|i \omega)$, has the following non-zero matrix elements at non-zero frequencies\cite{Maha}

\begin{align}
 T_{xx}^n(i\xi_n ) =T_{yy}^n(i\xi_n ) &=  \bigg[\bigg(\frac{\xi_n}{ c }\bigg)^2 + \bigg(\frac{\xi_n}{ c }\bigg) \dfrac{1}{\rho} + \dfrac{1}{{{\rho ^2}}}\bigg]\, \frac{{e^{ - \xi_n \rho /c}}}{\rho \, \epsilon_w(i \xi_n )},\label{Eq61}\\
T_{zz}^n(i\xi_n ) &=- 2\bigg[\bigg(\frac{\xi_n}{ c }\bigg) \dfrac{1}{\rho}+ \frac{1}{ \rho ^2 } \bigg] \frac{{{e^{ - \xi_n \rho  /c}}}}{{\rho  \, \epsilon_w(i \xi_n ) }}.
\end{align}

Here $c=c_0/\sqrt{\epsilon_w(i\xi_n )}$, where $c_0$ is the velocity of light in free space. Additionally, in the electrolyte at zero frequency, the corresponding non-zero matrix elements are\cite{Maha}

\begin{align}
T_{xx}^0(0) =T_{yy}^0(0 ) =  \bigg(\frac{\kappa_{\rm D}}{\rho}  + \frac{1}{\rho ^2}\bigg)\frac{e^{ - \kappa_{\rm D} \rho }}{\rho\,  \epsilon_w (0) },\\
T_{zz}^0(0 ) =-2\bigg( \frac{\kappa_{\rm D}^2}{2} + \frac{\kappa_{\rm D}}{\rho}+ \frac{1}{\rho^2 }\bigg)\frac{e^{ -\kappa_{\rm D} \rho}}{\rho \, \epsilon_w (0)  }\, ,
\end{align}
where $ \epsilon_w (0)$ is  the dielectric constant of water, and the inverse Debye-length in a monovalent electrolyte is

\begin{equation}
\kappa_{\rm D}=\frac{1}{\lambda_{\rm D}}=\sqrt{\frac{8 \pi N e^2}{\epsilon_w (0) k_B T}},
\label{Eq7b}
\end{equation}
where $N$ is the ionic concentration.

As noted above, the dielectric functions for electrolytes at non-zero frequencies are well approximated by water, but we will test what asymptotes follow when a simple model for the dielectric function is used and a temperature close to 0 K is considered. For this case ($\xi>0$), we use the following approximate dielectric function for an electrolyte from Davies and Ninham\,\cite{Davies}:

\begin{align}
\epsilon_w (i \xi_n )&=\epsilon_0 \left(1+\frac{\omega_{p}^2}{\xi (\xi+\eta)}\right)    
\end{align}
where the limits for low and high-frequency read
\begin{align}
 \lim_{\xi \rightarrow 0} \epsilon_w (i \xi_n )&=\frac{\epsilon_0 \omega_{p}^2}{\xi \eta} \label{Eq9} \\
\lim_{\omega_{p}/\xi \rightarrow 0} \epsilon_w (i \xi_n )&=\epsilon_0 \label{Eq 10} 
\end{align}

In free space at the non-retarded limit, the resonance interaction energy between two atoms in an orientationally averaged excited final state is zero. However, this is no longer true in an electrolyte. Here, there will be a contribution from the classical zero-frequency term,

\begin{align}
    U(\rho)= -k_B T\bigg( \alpha(0)^* \kappa_{\rm D}^2\bigg) \frac{e^{-\kappa_{\rm D} \rho}}{\rho \epsilon_w (0)}.
\end{align}

This term vanishes for zero temperature and also in free space. The fully retarded resonance interaction energy between two atoms in an orientationally averaged excited state at zero temperature is

\begin{equation}
U(\rho)=-{\frac{\hbar }{ \rho \pi c_0^2}}  \int_0 ^\infty d\xi \,\xi^2\, \alpha(i \xi )^*\,   e^{-\xi\,   \rho/c}.
\label{Eq12}
\end{equation}

For small separations, high frequencies (in relation to the plasma frequency so that Eq.\,\ref{Eq 10} applies) are important, and the following approximate expression may apply.

\begin{equation}
U(\rho)=-\frac{\hbar\, c_0}{\rho^4 \, \pi\, \epsilon_0^{3/2}}  \int_0 ^\infty dx x^2 e^{-x} \alpha \bigg(i x \frac{c}{\rho}\bigg)^*.
\label{Eq13}
\end{equation}

If the excess polarizability is approximated with its static limit, we find the $1/\rho^4$ dependence observed in free space. For large separations, the opposite applies: only the low frequencies contribute, then we get (using Eq. \ref{Eq9}) the asymptotic expression.

\begin{align}
U(\rho)=-\frac{2\hbar c_0^4 \, \alpha(0)^*}{ \rho^7 \pi \epsilon_0^3\, \omega_p^6/\eta^3} \, \int_0 ^\infty dx \,x^5\,   e^{- x}.
\end{align}
Here, we can more safely approximate the excess polarizability with the static limit. The long-range asymptote in an electrolyte at zero temperature decreases proportionally to $1/\rho^7$. 

\section{Summary and Discussions on Biological Impact}

The focus of this work was to demonstrate the need for more advanced basic and applied research into the non-local effects of salt solutions on dispersion forces. We investigated resonance interactions in the non-retarded limit. An important outcome of this study is the prediction of a crossover from an approximate power law $1/\rho^4$ to a $1/\rho^7$ dependence for excited atom pairs resulting from a retardation effect previously unpredicted in electrolytes.

It is important to note that this specific crossover is observed in the limit of zero or comparatively low temperatures. Notably, biological systems operate in liquid electrolytes at near-room temperatures. Under these realistic conditions, the interaction landscape changes fundamentally: a static contribution dominates, producing a screened separation dependence proportional to $e^{-\kappa_{\rm D}\rho}/\rho$. Furthermore, dissipation in the fluid medium is expected to significantly reduce the lifetime of excited states, resulting in these short-range and short-lived interactions.

Despite these damping effects, the sensitivity of van der Waals forces to background electrolytes raises critical questions about their influence on biological processes. As a concrete example of how non-local media impact these forces, we examined excited-state dipole-dipole interactions in the context of pheromone recognition in background media. While the biochemical consequences of pheromone detection are well understood, the initial recognition mechanism remains elusive. Based on an idea originally proposed by Dr. Zoltan Blum\,\cite{Ninhb}, it is suggested that resonance energy transfer in air, even when short-lived in non-local media, may be involved in how specific pheromone molecules activate receptor proteins in insect antennae. One purpose of this article was to outline a path toward a theoretical model for this recognition process. Ultimately, this work highlights the necessity for further theoretical development regarding ground- and excited-state interactions for atoms, molecules, and ions in non-local media.

\section*{Acknowledgments}
This research is part of the project No. 2022/47/P/ST3/01236 co-funded by the National Science Centre and the European Union's Horizon 2020 research and innovation programme               under the Marie Sk{\l}odowska-Curie grant agreement No. 945339. Institutional and infrastructural support for the ENSEMBLE3 Centre of Excellence was provided through the ENSEMBLE3 project (MAB/2020/14) delivered within the Foundation for Polish Science International Research Agenda Programme and co-financed by the European Regional Development Fund and the Horizon 2020 Teaming for Excellence initiative (Grant Agreement No. 857543), as well as the Ministry of Education and Science initiative “Support for Centres of Excellence in Poland under Horizon 2020” (MEiN/2023/DIR/3797).


\bibliographystyle{elsarticle-num} 
\bibliography{Larissa}

\end{document}